\documentclass{article}
\usepackage{PRIMEarxiv}
\usepackage[utf8]{inputenc} 
\usepackage[T1]{fontenc}    
\usepackage{lmodern}        
\usepackage{hyperref}       
\usepackage{url}            
\usepackage{booktabs}       
\usepackage{multirow}       
\usepackage{amsfonts}       
\usepackage{nicefrac}       
\usepackage{microtype}      
\usepackage{lipsum}
\usepackage{fancyhdr}       
\usepackage{graphicx}       
\usepackage{tikz}           
\usetikzlibrary{arrows.meta, positioning, fit, backgrounds, calc}
\graphicspath{{media/}}     

\usepackage{titlesec}
\titlespacing*{\section}
{0pt}        
{1.2ex}      
{0.8ex}      

\title{Open Security Benchmark: \\ Towards Autonomous Enterprise Cyber Defense}

\author{
  \textbf{Gal Engelberg},
  \textbf{Michael Arenzon},
  \textbf{Leon Goldberg}\\
  open.security (by Sola Security), Tel Aviv, Israel \\
  \texttt{research@open.security}
}

\begin{document}
\maketitle

\begin{abstract}
Enterprises are moving toward \emph{autonomous cyber defense}: agentic AI that builds situational awareness of an organization's security state and reasons from it to assessments, decisions, and actions. This rests on a \emph{holistic view of the enterprise's security state}, the continuous, cross-vendor picture of identities, cloud and infrastructure, data, applications, and their configurations that security posture management assembles. As agents take on this work, what matters is not whether an agent \emph{can} produce an answer but whether it \emph{should be trusted} to. The field cannot yet answer this question. Real enterprise environments are private, cross-vendor, and deeply correlated, and none is exposed publicly as a shared, queryable target for evaluating such agents end to end. We call this the \emph{environment data gap}. We present \textbf{Open Security Benchmark (OSB)}, a framework that benchmarks agentic AI on this work. OSB surfaces a curated enterprise environment-a frozen, holistic view of the security state-and evaluates posture investigation across two modalities: \emph{text-to-SQL} over a relational snapshot and each vendor's \emph{native API} over a served instance of the same environment. Freezing the environment pins the target state as an immutable snapshot and anchors answers to a closed-form ground truth. OSB is built from five components: a data layer, a task and evaluation-set layer, a multi-dimensional scoring layer, a minimal auditable harness, and a bring-your-own path that serves public comparison and private tenant evaluation from one substrate. We instantiate the framework with two identity-security packs and a family of synthetic-organization environment datasets spanning multiple scales, and chart its extension to further posture subdomains, investigation modalities, and defense stages from assessment toward remediation.
\end{abstract}


\section{Introduction}\label{sec:intro}
Autonomous cyber defense, systems that sense, reason about, and act on cyber threats with limited human intervention, is now an explicit goal in enterprise security, pursued across reviews of automated and autonomous cyber defense~\cite{Vyas2023_AutomatedCyberDefence} and training environments for autonomous agents~\cite{Standen2021_CybORG}. Yet every stage of the sense-reason-act loop presupposes one foundation: an accurate, current, holistic understanding of the organization's own security state. Situational-awareness research makes this precise~\cite{Endsley1995_SituationAwareness}, adapted in security as \emph{cyber situational awareness}~\cite{Franke2014_CyberSituationalAwareness}: an agent cannot act on what it cannot see.

Establishing this self-knowledge is therefore a prerequisite for autonomous cyber defense. Security posture management, the continuous, holistic assessment of an organization's defensive readiness and security state, supplies this situational-awareness foundation, on which any autonomous assessment or decision rests.

This assessment spans multiple subdomains: identity and access~\cite{RSA2025_ISPM}, cloud and infrastructure configuration~\cite{csa_ccm}, data exposure and classification~\cite{nist_sp_800_60}, application security and vulnerability state~\cite{owasp_top10}, third-party and SaaS integrations~\cite{nist_sp_800_161}, and compliance alignment with regulatory and contractual obligations~\cite{nist_csf2}. Catalogs like NIST SP~800-53~\cite{nist_sp_800_53r5}, the CIS Critical Security Controls~\cite{cis_controls_v8}, and ISO/IEC~27001~\cite{iso_iec_27001_2022} codify this work, and posture tooling from cloud auditors~\cite{scoutsuite, prowler} to SaaS scanners~\cite{scubagoggles} operationalizes it across vendor stacks.

Enterprises increasingly apply agentic AI to this work, with reported gains in coverage and analyst throughput~\cite{GoogleCloud2025_ROI_AI_Security}. As deployment scales, the binding question changes: an agent can produce an answer, but should we trust it, in this environment, with this data, given the evidence it produces and the ways it is known to fail? Answering this requires systematic, reproducible evaluation of agents on posture work, faithful to how it is done.

Recent cybersecurity benchmarks advance LLM evaluation across adjacent capabilities~\cite{ExCyTInBench_SecRL, OrgAccess_Benchmark_RBAC_Reasoning, CyberSecEval2}, but each addresses a narrow slice in isolation (\S\ref{sec:related}). More fundamentally, the field faces an \emph{environment data gap}: enterprise environments are proprietary, span many vendors, and interlink their data, and while executable environments exist for \emph{offensive} security tasks~\cite{Zhang2025_Cybench, Shao2024_NYUCTF}, none surfaces a cross-vendor enterprise environment as a shared, queryable substrate for evaluating \emph{defensive} posture-investigation agents end to end with gold posture answers. Practitioners therefore cannot compare agents on the work they must deploy them for, and researchers lack a stable target.

We present \textbf{Open Security Benchmark (OSB)}, a framework for evaluating agentic AI on security posture tasks \footnote{OSB is open source at \url{https://github.com/OpenSecurityAI/benchmark}.}. OSB closes the environment data gap by surfacing curated enterprise environments as a single, end-to-end view of the security state. Its methodological commitment is a frozen environment that agents investigate through the surface they would use in practice: \emph{text-to-SQL} over the relational snapshot, or each vendor's \emph{native API} over a served instance of the same environment. OSB aims at the full autonomous-cyber-defense loop, up through automated remediation, but starts at the assessment stage. Assessment is the read-only work of surfacing and substantiating findings, on which every later stage builds (\S\ref{sec:discussion}).

We formulate posture investigation as an agentic, execution-grounded text-to-SQL task. Each instance is a pair $(D, q)$: $D$ is a read-only relational snapshot of the organization's holistic security state (identities, cloud and infrastructure, identity providers, SaaS), and $q$ is a natural-language security question. The agent must discover the tables in $D$ relevant to $q$, compose read-only SQL queries that retrieve the supporting evidence, and synthesize a natural-language answer $a$. The answer is graded against a closed-form reference (a verdict, a closed set of matching subjects, and the determining configuration facts), so scoring is plan-independent: two semantically equivalent queries score alike, in the denotation-based tradition of semantic parsing that judges a parse by its answer, not its logical form~\cite{Berant2013_SemanticParsing}. This extends the schema-realistic, agentic text-to-SQL paradigm of the Spider and BIRD lineage~\cite{yu2018spider, Li2023_BIRD, Lei2024_Spider2} to defensive security work, casting each investigation as a reason-and-act loop~\cite{Yao2023_ReAct} over a frozen environment behind a constrained tool interface~\cite{Yang2024_SWEagent}. A second modality keeps the same $D$, question, and answer criteria but changes the surface: the agent interrogates a served instance of the same environment over its vendors' native APIs with real vendor tooling, and its recorded request trace stands in for the SQL the relational modality would produce (\S\ref{sec:surface}).

We ground this general substrate in a concrete first domain. Identity security posture management (ISPM) is a natural start: cross-vendor by nature, data-rich, consequential, and newly equipped with public benchmark definitions we build on~\cite{SolaISPMVisibility2026, SolaISPMCrossVendor2026}. The framework extends across posture subdomains under the same methodology. It also extends along the progression that autonomous cyber defense automates: assessment, risk assessment and prioritization~\cite{nist_sp_800_30}, mitigation and remediation recommendation~\cite{nist_sp_800_40}, remediation and response implementation~\cite{Vyas2023_AutomatedCyberDefence, Standen2021_CybORG}, and post-remediation verification via continuous monitoring~\cite{nist_sp_800_137}. It instantiates only the assessment stage (\S\ref{sec:discussion}).

This paper makes three contributions: (i) \textbf{Open Security Benchmark}, an evaluation framework whose five components (a data layer, a task and evaluation-set layer, a multi-dimensional scoring layer, a minimal auditable harness, and a bring-your-own path) compose a standardized substrate for both relational and native-API investigation, with an explicit trust boundary; (ii) a curated \emph{data catalog} of multi-vendor enterprise environments, packaged as synthetic-organization datasets that close the environment data gap; and (iii) initial \emph{use-case packs} (ISPM~Visibility and ISPM~Cross-Vendor) that instantiate the framework end to end and open an extension path for community-authored packs across posture subdomains.

The rest of the paper is structured as follows. Section~\ref{sec:related} surveys related work and positions OSB against it. Section~\ref{sec:surface} presents the two investigation modalities and why text-to-SQL is standardized first. Section~\ref{sec:components} describes the framework's five components. Section~\ref{sec:data} presents the environment datasets and Section~\ref{sec:packs} the benchmarking packs. Section~\ref{sec:discussion} discusses broader implications and limitations, and Section~\ref{sec:conclusion} concludes.

\section{Related Work}\label{sec:related}
AI evaluation in security and data systems has expanded to measure models' reasoning, operational reliability, and alignment with real-world investigative workflows. Three threads dominate: text-to-SQL benchmarks that assess data-centric reasoning; cybersecurity benchmarks spanning SOC, CTI, vulnerability, and access-governance reasoning; and methodological work on automated evaluation. Open Security Benchmark draws on all three.

The Spider benchmarks are the canonical foundation for natural-language database interfaces. Spider~1.0~\cite{yu2018spider} introduced cross-domain text-to-SQL over a static schema-query paradigm, scored by exact set match and, later, execution accuracy. BIRD~\cite{Li2023_BIRD} moved toward large, ``dirty'' databases demanding external-knowledge grounding and value semantics, close to the messy-enterprise character OSB targets. Spider~2.0~\cite{Lei2024_Spider2} embeds models in enterprise environments such as BigQuery and Snowflake, requiring multi-step workflows, error revision, and reasoning over code-schema dependencies. Together these mark a transition from scoring a single SQL output to evaluating the robustness and repairability of agentic workflows, the transition OSB follows for defensive security investigation.

ExCyTIn-Bench~\cite{ExCyTInBench_SecRL} rewards step-by-step investigation in a simulated SOC aligned with a threat investigation graph, while CyberSOCEval~\cite{CyberSOCEval_CrowdStrike} grounds evaluation in authentic threat reports and sandbox outputs. For cyber threat intelligence, CTIBench~\cite{CTIBench} mixes expert-validated answers with CVE/CWE mappings, and SEvenLLM-Bench~\cite{SEvenLLM_Bench} adds a bilingual setting with hybrid semantic scoring. SecLLMHolmes~\cite{SecLLMHolmes_AI4CloudOps} benchmarks vulnerability detection over labeled C/C++ and Python code. General-purpose suites add breadth: CyberBench~\cite{CyberBench_JPMC} spans classification, summarization, NER, and QA; CyberMetric~\cite{CyberMetric} measures certification-level knowledge; and CyberSecEval~2~\cite{CyberSecEval2} evaluates insecure code generation, prompt-injection susceptibility, and code-interpreter abuse.

Identity and access reasoning, central to modern posture, gained attention through OrgAccess~\cite{OrgAccess_Benchmark_RBAC_Reasoning}, which builds synthetic hierarchies, permission sets, and RBAC rules to test whether models apply access-governance policies. Where OrgAccess tests RBAC application within synthetic structures, OSB targets operational visibility in realistic environments: enumerating identities, identifying misconfigurations, and evaluating authentication posture over production-shaped IAM, identity-provider (IdP), and SaaS data, within a substrate that generalizes beyond identity to other posture subdomains. The two Sola ISPM benchmarks~\cite{SolaISPMVisibility2026, SolaISPMCrossVendor2026}, which OSB packages as its first packs, established the single-platform visibility and cross-vendor correlation task families on which we build.

Two adjacent lines make OSB's gap concrete. First, executable \emph{agent} environments have appeared for \emph{offensive} security: Cybench~\cite{Zhang2025_Cybench} and, at scale, NYU~CTF~Bench~\cite{Shao2024_NYUCTF} run agents over capture-the-flag tasks, and vulnerable-by-design cloud labs ship shareable misconfigured environments~\cite{cloudgoat}. These are shared, executable substrates, but they target attack rather than defensive posture assessment and carry no gold posture answers for end-to-end scoring. Second, configuration-as-SQL tooling such as Steampipe~\cite{steampipe} and CloudQuery~\cite{cloudquery} already exposes cloud, IdP, and SaaS \emph{configuration state} as queryable relational tables, direct evidence that posture-as-SQL is a natural, increasingly common representation. But these are query engines and human-run audit tools, not agent benchmarks: they provide neither a fixed task suite with gold answers nor a scoring methodology for comparing agents. What OSB provides is the missing piece: a shared, defensively oriented, cross-vendor relational environment paired with gold posture answers and a reproducible scoring methodology for end-to-end agent evaluation.

A complementary line addresses automated-evaluation methodology. G-Eval~\cite{Liu2023_GEVAL} uses LLMs as evaluators with chain-of-thought form-filling prompts that align well with human judgment; RAGAS~\cite{Es2023_RAGAS} introduces reference-free metrics for context relevance and answer faithfulness. These inform how OSB constructs scoring: deterministic checks where exact agreement is meaningful, rubric-locked LLM judges where exact matching breaks down, and expert calibration to keep both honest.

OSB draws on this trajectory in three ways: it adopts text-to-SQL over a frozen relational snapshot as a unifying paradigm, extending Spider and BIRD to defensive workflows; it integrates deterministic structural metrics, rubric-locked LLM-judge panels, and expert calibration as distinct, attributable axes rather than one aggregate; and it treats public comparability and private tenant evaluation as core properties via a shared substrate and explicit trust boundary. Where each prior benchmark addresses a narrow capability slice, OSB is a framework into which packs spanning identity, configuration, vulnerability, and broader posture subdomains plug under a shared methodology.

\section{Investigation Modalities}\label{sec:surface}
Posture review is investigative: correlating data across many sources into a holistic view of the enterprise's security state, then reasoning over it to surface and substantiate a finding. This situational awareness is the foundation of autonomous enterprise cyber defense, and an agent doing this work must \emph{act} on the environment through some interrogation surface. OSB instantiates two \emph{investigation modalities} over the same environment: relational queries and native vendor interfaces. An agent is thus measured on the surface it would actually use, while the environment, tasks, and answer criteria stay shared.

An agent can interrogate an enterprise environment along at least three surfaces. \emph{Relational queries} (text-to-SQL over a centralized store, where each pivoted entity becomes a table, each cross-vendor relationship a join key, and each posture question a workload over that schema) are the surface studied by the Spider lineage~\cite{yu2018spider, Lei2024_Spider2}. \emph{API and tool calls} against live vendor endpoints are exercised by tool-using agents~\cite{Yao2023_ReAct} and measured by API- and function-calling benchmarks~\cite{Patil2023_Gorilla, Li2023_APIBank}. \emph{Shell and command-line invocation} of vendor CLIs in a sandbox is studied under agent--computer interfaces~\cite{Yang2024_SWEagent}. Each route reaches the same finding but defines ``done the work'' differently; OSB implements the relational surface and the native-vendor surface, the latter spanning both direct API calls and vendor command-line tools.

The relational modality is faithful to the investigative activity: posture analysts \emph{correlate} across sources (joining, pivoting, filtering), and a relational query is a direct, legible encoding of that correlation. It is also, decisively for a benchmark, the surface whose reproducible target is cheapest to obtain and check, because posture \emph{state} is naturally a relational snapshot that freezes immutably and admits cheap set-equivalence checking. Reproducibility comes from \emph{freezing or recording} the target, not from SQL as such: API-Bank drives simulated API executors, Gorilla scores against a static API database, and SWE-bench-style harnesses run agents in frozen container images with deterministic tests~\cite{Li2023_APIBank, Patil2023_Gorilla, Yang2024_SWEagent}. Text-to-SQL is effective precisely when posture-relevant configuration and identity state, not security telemetry, is available as centralized, normalized, queryable relational data: configuration-as-SQL engines such as Steampipe~\cite{steampipe} and CloudQuery~\cite{cloudquery} already present cross-vendor configuration state as relational tables, whereas the Open Cybersecurity Schema Framework~\cite{ocsf} and managed security lakes~\cite{aws_security_lake} normalize \emph{telemetry} rather than the inventory and entitlement state OSB queries. The hard, often unsolved part is \emph{cross-vendor identity resolution}: reconciling that a person in an HR system, an identity provider, and several clouds is one identity, which these engines rarely expose as clean foreign keys. OSB therefore keeps it in the task rather than assuming it solved (\S\ref{sec:packs}).

The native-vendor modality measures the work as an operator actually does it, for the many settings where no consolidating relational layer exists. The same environment snapshot is compiled into a running world that serves the enterprise over its vendors' \emph{native} API surfaces; the agent interrogates it with real vendor tooling: the \texttt{aws} CLI and ordinary HTTP calls to the identity provider, productivity suite, source host, and HR system, authenticating as an operator would. Its only documentation is the pinned, real vendor API specifications, and it never sees the underlying tables: it observes the environment solely through computed, deterministic, read-only API responses, so reproducibility follows from the frozen snapshot behind a faithful emulator rather than from freezing a query result. Every request the agent issues is recorded to a request log that plays the role the query trace plays for the relational modality-the evidence the scorer reads. The native-vendor modality trades cheap set-equivalence for fidelity to how the work is really done, and needs no pre-consolidated schema.

Both modalities share the scoring layer, the environment, and the answer criteria; they differ only in the interrogation surface and the trace it produces. Text-to-SQL is standardized first because its verifiable target is cheapest, and the metrics that inspect the \emph{query artifact}-SQL quality and the structural table and join checks (\S\ref{sec:metrics})-are specific to it, while answer correctness and reasoning quality apply to both. Neither surface is a ceiling: because scoring is defined independently of the surface, further modalities (a graph surface for transitive access paths, cloud-posture-audit checks) plug in under the same methodology (\S\ref{sec:discussion}).

\section{Framework Components}\label{sec:components}
Open Security Benchmark comprises five components that together implement the commitment argued in Section~\ref{sec:surface}. A \emph{data layer} (\S\ref{sec:datalayer}) surfaces the enterprise environment as a frozen, shared snapshot. A \emph{task and evaluation-set layer} (\S\ref{sec:tasks}) defines security questions and their expected structured answers. A \emph{scoring layer} (\S\ref{sec:metrics}) reports each criterion separately. A \emph{harness layer} (\S\ref{sec:harness}) provides a small, fixed agent action space that makes every run auditable and reproducible. A \emph{bring-your-own path} (\S\ref{sec:byo}) lets researchers and practitioners use the framework from one substrate without compromising one another's needs. This section describes each component at the level of design intent; concrete instantiations appear in Sections~\ref{sec:data} and~\ref{sec:packs}.

\subsection{Enterprise Environment Data}\label{sec:datalayer}
Posture review runs against the operational state of an enterprise: identities, resources, role and group memberships, configuration settings, vulnerabilities, and the access paths that connect them. OSB's data layer exposes this state as a curated relational database, under the entity-to-table mapping of Section~\ref{sec:surface} and for the reasons argued there.

Vendor coverage spans the layers where posture risk concentrates (cloud and infrastructure, identity providers, SaaS suites, application and dependency inventories, data stores); the present packs instantiate a subset. Within each vendor, the schema captures the entities and relationships that rule sets such as Scout Suite~\cite{scoutsuite} and ScubaGoggles~\cite{scubagoggles} interrogate, plus the artifacts vendor best-practice guidance~\cite{okta_security_best_practices, gws_admin_security} treats as audit-relevant. Across vendors, the \emph{joinable columns} on which multi-stack reasoning depends-shared emails and usernames, group and resource identifiers, ownership fields-are documented but deliberately \emph{not} wired as declared foreign keys, so the agent must infer which columns correspond and cross-vendor identity resolution stays part of the task (\S\ref{sec:packs}). This read-only relational data is the shared foundation, without precluding other substrates (log streams, graph telemetry, simulated endpoints); transitive, path-based questions (nested group membership, privilege-escalation chains) are weakest in flat SQL, so a graph surface in the spirit of identity attack-path tooling~\cite{bloodhound} may later serve them better (\S\ref{sec:discussion}).

A trust boundary separates what the agent observes (the schema and its contents) from what only the evaluator observes (the gold reference, the expected result, and the seed data binding). Datasets are produced by a generator pipeline that cross-checks each question against the data before release, but that pipeline sits outside the benchmark: the shipped artifact is the frozen snapshot plus the evaluator-only reference, not an authoring trace. Because the same contract holds whatever organization the data describes, a benchmark authored against one environment ports to another without re-engineering the methodology.

\subsection{Tasks and Evaluation Sets}\label{sec:tasks}
A task pairs a natural-language security question with a SQL-derivable structured answer (the $(D,q)$ formulation of \S\ref{sec:intro}). That answer is either a list of entities with the fields an operator consumes or a posture verdict with the configuration facts behind it; for an enumeration question, the answer is the set of matching subjects, empty when none qualify. Because the answer is specified independently of the query that derives it, equivalent queries receive the same score. On the evaluator side, each task carries a gold reference that produces the expected answer against the same data the agent sees, and the trust boundary keeps this reference in a separate visibility class, accessible only to the scorer.

An \emph{evaluation set} is a coherent collection of tasks scoped to a domain or capability, such as identity hygiene, configuration drift, or cross-vendor reasoning. The contract is designed around three visibility classes: an open class for public comparison and ablation, a gated class that withholds the reference artifacts to protect leaderboard integrity, and a private class that keeps tasks and evaluators inside a tenant's boundary. Open comparison and private bring-your-own evaluation ship today; the gated leaderboard channel is on the roadmap (\S\ref{sec:discussion}). The same task definition serves public comparison and internal review without duplication.

\subsection{Evaluation Metrics}\label{sec:metrics}
OSB scores every run using evaluation criteria adopted from the cross-vendor ISPM benchmark~\cite{SolaISPMCrossVendor2026}, reusing its Appendix~B rubric texts verbatim rather than re-deriving them. Scoring proceeds in a four-stage pipeline: execute every question, collect a per-question evidence bundle, apply a rubric-guided LLM-as-judge panel, and run a deterministic structural check. Judged metrics use a three-level scale ($0$, $0.5$, $1$), except the binary answer-verdict check. The judge design follows work on LLM-as-judge evaluation~\cite{Liu2023_GEVAL} and retrieval-augmented evaluation~\cite{Es2023_RAGAS}, applying rubric-locked panels with explicit thresholds rather than free-form scoring. The deterministic structural metrics add a verification layer immune to the semantic ambiguity of purely LLM-based judgment.

The framework avoids collapsing these criteria into a single aggregate: one number obscures the source of failure, whereas separate metrics let teams attribute performance shifts to specific behaviors and read a scorecard in the language of the property they care about. Answer correctness is the primary measure, graded against the answer fields of the task's ground-truth atom (\S\ref{sec:data}): a verdict, a closed set of matching subjects, and the determining configuration facts. Because the agent answers in free-form prose, it is scored by a rubric-locked judge that maps the answer onto that reference, absorbing benign variation such as how subjects are named or how a count is reported (``about 12'' versus an exact 12) rather than requiring naive string equality. The other judged criteria assess the yes/no verdict, the utility of the reasoning trace, and the schema-consistency of the generated SQL, while the deterministic structural metrics measure whether the agent constructed the required relational path, the tables and joins the finding needs. All are compared against the ground-truth atom or its minimal-sufficient evidence set (the smallest set of tables and joins that suffices to derive the answer) rather than a particular query plan, so scores are plan-independent (\S\ref{sec:tasks}); the criteria are complementary but not statistically independent: wrong tables tend to produce wrong answers, so we report the relationships among them rather than assume orthogonality. The answer-correctness, verdict, and reasoning criteria apply to both investigation modalities; the SQL-quality criterion and the structural table and join metrics are specific to the text-to-SQL modality, since they inspect the query artifact. In the native-API modality the same structural role is played by whether the agent's recorded request trace reaches the endpoints and objects the finding requires (\S\ref{sec:surface}).

\begin{table}[t]
\centering
\small
\caption{OSB's evaluation criteria for posture investigation, with rubric texts
adopted from the cross-vendor ISPM benchmark~\cite{SolaISPMCrossVendor2026}.}
\label{tab:metrics}
\footnotesize
\setlength{\tabcolsep}{4pt}
\begin{tabular}{@{}p{1.4cm}p{4.2cm}p{1.55cm}p{1.55cm}p{4.25cm}@{}}
\toprule
\textbf{Family} & \textbf{Criterion} & \textbf{Scale} & \textbf{Type} & \textbf{Description} \\
\midrule
\multirow{2}{1.4cm}{Answer quality} & AnswerCorrectnessVsGT & 0/.5/1 & judged & Answer agrees with the ground-truth verdict, subjects, and facts. \\
 & AnswerVerdictCorrectnessVsGT & 0/1 & judged & Primary yes/no verdict matches ground truth. \\
\addlinespace
Reasoning & ReasoningUtility & 0/.5/1 & judged & Reasoning trace makes plausible progress toward the answer. \\
\addlinespace
SQL quality & SQLSemanticAppropriateness & 0/.5/1 & judged & SQL is a sensible, schema-consistent strategy for the question. \\
\addlinespace
\multirow{2}{1.4cm}{Structural} & Tables/Joins Recall & [0,1] & det. & Share of the required tables/joins the agent used. \\
 & Tables/Joins Prec., F1 & [0,1] & det. & Precision and F1 over tables/joins. \\
\bottomrule
\end{tabular}
\end{table}

Beyond the per-metric scores, OSB breaks every metric down within a run across the task taxonomy (complexity, focus, platform, and single- versus cross-platform scope; \S\ref{sec:packs}), and across environment variants (organizational scale today, and the broader matrix of vendor stack, schema variant, and noise as the catalog grows; \S\ref{sec:data}) by comparing runs. Generalization and robustness are thus read from the same criteria across configurations rather than measured separately: an agent that exploits one configuration is distinguished from one that holds up across the matrix. This breakdown is a reporting practice, not an additional criterion.

Scores come from a panel of two judges from different providers, paired to reduce the correlated and self-preferring biases LLM judges are known to exhibit~\cite{Zheng2023_JudgeLLM, Wang2023_FairEval}. Each task is run for several independent traces; every trace is scored by every judge; and a metric's score is the mode of the pooled votes (majority voting over samples in the spirit of self-consistency~\cite{Wang2023_SelfConsistency}), ties broken toward the lower grade. Because frontier judges drift, scores are comparable only within a fixed judge configuration, and open-weights judges may be substituted for reproducibility or to keep a private tenant's data in-boundary (\S\ref{sec:byo}). To expose scorer noise rather than hide it in a point estimate, every metric is reported with a bootstrap 95\% confidence interval (resampling over tasks) and inter-judge agreement (Cohen's $\kappa$~\cite{Cohen1960_Kappa}). Finally, the framework treats expert calibration as its anchor: automated scores are to be validated against expert-labeled subsets, with their agreement reported so scorer drift stays visible.

\subsection{The Harness}\label{sec:harness}
The harness is the execution substrate that runs an agent against a task and produces the artifacts on which every metric is computed. A single principle governs its design: every action an agent takes must be observable and reproducible from the recorded run alone. This rules out hidden state, opaque tool calls, and free-form interaction patterns that cannot be replayed.

The agent acts through two tools. \textsc{get-schema} returns the schema documentation for the whole environment in a single view: every table with its columns, row counts, and a short note. It exposes column names but neither types nor foreign keys, so the agent must infer every join, within and across vendors, from the columns themselves (\S\ref{sec:datalayer}). \textsc{run-query} executes a single SQL statement and returns its result. Each task runs against a private, disposable copy of the snapshot, so although any statement is permitted, writes touch only that copy and the shared environment is never mutated: posture review is non-mutating by nature, and copy-isolation gives that guarantee without constraining the agent's SQL. The surface is small enough that agent behavior maps cleanly onto measurable events: query iteration is the trajectory of exploratory and candidate queries, and answer verification is the presence and structure of a validating query.

Beyond the tool surface, each run produces an \emph{evidence archive}: an ordered trace of every model step, tool call, and result, written to disk (\S\ref{sec:discussion}). The agent's reasoning is interleaved with this trace as the text between its tool calls, and a reasoning journal is recovered from it for scoring. Because reasoning and actions share one ordered record, any claim a scorer makes is checkable against it: a reasoning-utility metric that flags an unsupported assertion can point to the surrounding tool calls, and a structural metric that detects a wrong join can point to the query in which the join appears.

The harness is the only component the framework standardizes at the implementation level; the rest of the agent runtime (model serving, prompting strategy, orchestration loop) is the experimental variable. The framework already runs several agent runtimes over this one tool interface, and any model can be placed behind it, so cross-stack comparability follows from pinning what every agent must produce rather than how it is built.

\subsection{Bring Your Own Agent or Model}\label{sec:byo}
The components above compose into one path for both researchers and practitioners. A researcher evaluating an agent or model points one of the framework's supported runtimes at a published pack, runs it under the standard harness, and receives a scorecard across all criteria together with the full evidence archive. The pack is version-pinned and its evaluator artifacts are gated, so scorecards are comparable across runs on the same pack. An ablation is a re-run against that pinned pack with a single variable changed, whether the model, the prompt strategy, or the context configuration, and the resulting metric movements are attributable to it. Comparability comes from pinning the artifacts those choices must produce, not from enforcing a specific model or runtime.

A practitioner evaluating an agent against a specific tenant follows a structurally identical path, with the same harness, contract, and criteria, but supplies a private pack: tenant-specific data and the success definitions (the expected answers) for the benchmark's questions. Because the judged metrics send the agent's evidence bundle to an LLM panel, a private pack can run that panel on self-hosted or open-weights judges \emph{inside} the tenant boundary, so confidential posture data never leaves it, while the deterministic metrics require no external call at all. The metric layer accepts additional rubric-defined criteria for such packs (procedure alignment, control coverage, internal risk definitions), and the trust boundary keeps any private evaluator's outputs out of public comparison, so private evaluation can be as opinionated as a tenant needs without affecting the public surface.

\section{Environment Datasets}\label{sec:data}
The data catalog closes the environment data gap with \emph{environment datasets of synthetic organizations}: each a fully fictional enterprise with no real individuals, organizations, or credentials, every identifier (names, emails, logins, keys) fabricated on \texttt{.example} domains. Although every value is fabricated, the environments are synthesized from real ones: each vendor's tables and columns follow that product's real data model, and their data patterns (identity lifecycles, group and privilege structures, common misconfigurations, credential ages, and access distributions) mirror those seen in production enterprises. Tasks therefore exercise realistic posture reasoning while the snapshot exposes no real identity, credential, or organization. The datasets will be released under the Open Security AI organization on the Hugging Face Hub,\footnote{\url{https://huggingface.co/OpenSecurityAI}} each an immutable, content-addressed revision that pins exactly the bytes a run was reviewed against; the initial catalog is in preparation, and the counts below describe the reference snapshot.

\subsection{Relational encoding for text-to-SQL}
Each environment is a single read-only relational snapshot of an organization's identity-relevant state across eight vendor-style sources-AWS (IAM and Identity Center), Azure Active Directory, Google Cloud Platform, GitHub, Google Workspace (with Drive), a human-resources information system, MongoDB Atlas, and Okta-comprising 44 relational tables in the reference release. Every vendor entity and its configuration lives in a table (Table~\ref{tab:schema}); vendors are linked by \emph{joinable columns}, shared identifiers documented in the schema but, per \S\ref{sec:datalayer}, deliberately not declared as foreign keys, so cross-vendor identity resolution remains part of the task. A shared email, for instance, links \texttt{hibob\_employee} and \texttt{okta\_user} to the same person in \texttt{aws\_identitystore\_user}, so a question like ``terminated employees who still have cloud access'' becomes a join over those tables filtered by \texttt{hibob\_employee.status}. The two packs of \S\ref{sec:packs} contribute 127 tasks, realized at every scale.

\begin{table}[t]
\centering
\footnotesize
\setlength{\tabcolsep}{5pt}
\caption{Illustrative slice of the relational encoding.}
\label{tab:schema}
\begin{tabular}{@{}p{2.3cm}p{4.5cm}p{7.0cm}@{}}
\toprule
\textbf{Vendor source} & \textbf{Example table} & \textbf{Representative columns} \\
\midrule
\multirow{3}{*}{AWS}
  & \texttt{aws\_iam\_user}
  & \texttt{name, mfa\_enabled, password\_enabled, password\_last\_used} \\
  & \texttt{aws\_iam\_access\_key}
  & \texttt{access\_key\_id, user\_name, status, create\_date, last\_used\_date} \\
  & \texttt{aws\_iam\_account\_\allowbreak password\_policy}
  & \texttt{minimum\_password\_length, max\_password\_age, password\_reuse\_prevention} \\
\addlinespace
\multirow{2}{*}{Okta}
  & \texttt{okta\_user}
  & \texttt{login, email, status, last\_login, department} \\
  & \texttt{okta\_mfa\_policy}
  & \texttt{name, status, authenticators} \\
\addlinespace
Google Workspace
  & \texttt{googleworkspace\_\allowbreak shared\_drive\_file}
  & \texttt{name, owner\_email, permissions, trashed} \\
\addlinespace
HRIS (HiBob)
  & \texttt{hibob\_employee}
  & \texttt{email, fullname, status, lifecyclestatus, terminationdate} \\
\addlinespace
AWS Identity Center
  & \texttt{aws\_identitystore\_user}
  & \texttt{id, name, display\_name} \\
\bottomrule
\end{tabular}
\end{table}

\begin{table}[t]
\centering
\small
\caption{The reference environment at three organizational scales.}
\label{tab:scale}
\begin{tabular}{@{}lrr@{}}
\toprule
\textbf{Scale} & \textbf{Employees} & \textbf{Total rows} \\
\midrule
Small & $\sim$75 & $\sim$1{,}600 \\
Mid & $\sim$400 & $\sim$8{,}300 \\
Large & $\sim$2{,}000 & $\sim$36{,}000 \\
\bottomrule
\end{tabular}
\end{table}

A single environment cannot test generalization, so the reference organization is realized at three scales (small, mid, and large; Table~\ref{tab:scale}), sharing task structure but spanning more than an order of magnitude in size, from $\sim$75 people and $\sim$1{,}600 rows to $\sim$2{,}000 people and $\sim$36{,}000 rows over the same 44 tables (roughly a 5x step between consecutive scales). This instantiates the \emph{scale} axis of the environment matrix, so generalization and robustness (\S\ref{sec:metrics}) are reported as coverage across scale rather than one number, distinguishing a scale-robust agent from one that exploits a small configuration's idiosyncrasies. The broader matrix (vendor stack, schema variant, noise profile) is built into the data-layer contract as the natural direction of growth.

Each dataset is a self-contained bundle: a manifest with the identity header, the SQLite snapshot, and a per-task ground-truth file. That file is a \emph{ground-truth atom}: the answer fields (a verdict, a closed set of matching subjects, and the determining configuration facts) plus the tables and joins the finding touches. Time-relative questions (``inactive for 90 or more days,'' ``left the company but still has access'') anchor to a single publish date in the manifest, so look-back semantics belong to the data, not wall-clock time. Because each revision is immutable and self-contained, sharing no rows or answers with its predecessor, an environment can be rotated: regenerated and republished. Rotation is the lever against \emph{answer-level} memorization, since memorizing one environment's rows and answers yields no advantage on its successor. It does not defend against overfitting to the fixed public schema, task set, and solution strategies, which persist across rotations and scales; held-out task splits and a held-out environment generator (future work, \S\ref{sec:discussion}) are the guard for that stronger notion.

\subsection{The same environment, served natively}
The relational snapshot is not the only way an agent meets this environment. For the native-API modality (\S\ref{sec:surface}), the same bundle-the manifest plus the identical SQLite snapshot-is compiled into a self-contained, runnable world that serves the enterprise over its vendors' \emph{native} API surfaces: the AWS control plane reached through the \texttt{aws} CLI, and the identity provider, productivity suite, source host, and HR system reached over their real HTTP APIs. Responses are computed from the same rows the SQL modality queries, so the two modalities share one ground truth; but the agent sees them only as vendor API results, authenticates with vendor-native credentials, and reads the pinned real vendor API specifications as its sole documentation, never seeing the tables. A subset of the eight sources is served this way today, and one content-addressed revision underlies both presentations, so a finding is defined once and can be posed either as a query workload or as a sequence of native API calls.

\section{Use-Case Packs}\label{sec:packs}
A \emph{pack} is the unit in which OSB is instantiated and extended: a data domain, an evaluation set, and the scoring configuration that binds them. The framework ships today with two identity-security packs, derived from the two public Sola ISPM benchmarks, over the environment datasets of Section~\ref{sec:data}. Together they comprise 127 tasks-77 in the visibility pack and 50 in the cross-vendor pack-each realized at all three environment scales.

The visibility pack, derived from the single-platform Sola Visibility ISPM benchmark~\cite{SolaISPMVisibility2026}, evaluates foundational identity inventory and configuration-hygiene questions scoped to one vendor at a time. Representative questions include \emph{``Which active access keys are older than 90 days?''} over AWS and \emph{``Which applications have \texttt{mfa\_required} set to false or null?''} over Okta (a synthetic-schema simplification; production Okta enforces multi-factor requirements through sign-on policies rather than a per-application flag). These are the questions a posture analyst asks first (who and what exists, and whether baseline hygiene controls are in place), and they establish whether an agent can enumerate identities and read configuration state correctly before any cross-vendor reasoning.

The cross-vendor pack, derived from the Cross-Vendor Sola ISPM benchmark~\cite{SolaISPMCrossVendor2026}, evaluates federated correlation across vendor boundaries, where no shared schema or explicit foreign key connects the sources and the agent must reconstruct identity resolution paths itself. Representative questions include \emph{``Which terminated employees still have AWS single sign-on access?''}, which correlates a human-resources record against cloud access, and \emph{``Which publicly shared Google Workspace files are owned by Okta users who have been deactivated?''}, which joins an identity provider against a SaaS suite's sharing state to surface an exposure set. These questions exercise the correlation that makes identity posture hard and that no single-vendor tool can answer.

Both packs share a lean task taxonomy that drives every reported breakdown: \emph{complexity} (easy, medium, hard); \emph{focus} (authentication, authorization, and administration, the last covering lifecycle and hygiene such as offboarding and dormant credentials); and the concrete \emph{platforms} a task's ground truth touches. From the platforms, a single- versus cross-platform \emph{scope} is derived. Reporting each metric across these dimensions turns a single score into a diagnosis: an agent may enumerate single-platform inventory well yet fail cross-platform correlation, and the taxonomy makes that visible. The pack is also the unit of community contribution: a new posture subdomain is a new domain and evaluation set composed with the existing scoring methodology, so the path from these two identity packs to packs for cloud configuration, vulnerability state, or data exposure is an authoring task, not a framework redesign.

\section{Discussion}\label{sec:discussion}
OSB is not a scoreboard-it is an engine that can both measure an agent and advance it toward autonomous enterprise cyber defense, and the mechanism for the second role is already present in the first. Every run writes an evidence archive to disk (\S\ref{sec:harness}): the agent's actions through the \textsc{get-schema} and \textsc{run-query} tools, the intermediate results those queries returned, the reasoning interleaved between them, and the final answer, together with the scorecard derived from that record. The archive is browsable through the framework's run viewer and stamped with the binary and dataset revision it was produced against, so a run can be reopened and read against the exact bytes it saw. Three caveats qualify it: the prompt is reconstructible from the pinned pack inputs rather than stored verbatim, a stored run can be reviewed again but not yet automatically re-scored, and very large tool outputs are truncated in the record. These are the same artifacts a training loop consumes. OSB implements no training loop, but the pipeline that scores an agent today produces the versioned, contract-conformant, replayable data one would use to train its successor.

One potential utility of OSB is therefore to advance agents, not only to score them: what makes the archive trainable rather than merely readable is that OSB already computes the signal a learner needs. An outcome the framework can check automatically-a matched subject set, a correct verdict-is a verifiable reward, the signal behind recent reasoning models~\cite{Guo2025_DeepSeekR1, Lambert2024_Tulu3}. Where the outcome is partial-credit or rubric-graded rather than exactly checkable, a reward model learned from preferences or feedback supplies the same signal~\cite{Ouyang2022_InstructGPT, Christiano2017_RLHF}. The logged trajectories are the input distribution for offline reinforcement learning~\cite{Levine2020_OfflineRL} and for imitation learning from the successful traces~\cite{Ross2011_DAgger}. Filtering the runs OSB has already verified as correct and fine-tuning an agent on them closes the loop as bootstrapped self-improvement~\cite{Zelikman2022_STaR, Xu2025_DRO_Reasoning}, with OSB's metric outputs entering these procedures as the scalar or vector reward. None of it is a separate system; it reuses the scorecard the evaluator already emits. Our work is nonetheless limited by its reliance on \emph{synthetic environments}: they are shareable, reproducible, and privacy-safe, and because each is synthesized from real product data models and the behavioral patterns of production enterprises (identity lifecycles, group and privilege structures, common misconfigurations), the posture reasoning they demand carries over to real deployments; what a fabricated organization does not reproduce is a live tenant's operational drift and incident telemetry.

Future work includes growing the catalog and extending the surface. Because a shared benchmark advances a field only while it resists contamination~\cite{Deng2009_ImageNet, Wang2019_SuperGLUE, Jimenez2024_SWEbench}, we will publish more synthetic organizations and use-case packs, broadening coverage across schemas, vendor stacks, and enterprise archetypes while hardening the target as current packs saturate~\cite{Wang2019_SuperGLUE}. The scoring core already spans text-to-SQL and the native-vendor modality, and further surfaces plug into it the same way: cloud-posture audits after CIS~\cite{cis_controls_v8}, Prowler~\cite{prowler}, and Scout Suite~\cite{scoutsuite}, a graph surface for the transitive access-path questions that flat SQL handles awkwardly~\cite{bloodhound}, and new packs for configuration, vulnerability, and data-exposure posture. Above all, because autonomy is graded rather than binary~\cite{Parasuraman2000_Automation}, the read-only assessment OSB scores today-grounded in situation awareness~\cite{Endsley1995_SituationAwareness}-extends up the ladder toward risk prioritization, remediation, and verification, the goal of autonomous cyber defense~\cite{Vyas2023_AutomatedCyberDefence, Standen2021_CybORG}, with the evaluation-to-training loop above carrying an agent from each rung to the next.

\section{Conclusion}\label{sec:conclusion}
We introduced Open Security Benchmark, a framework for evaluating agentic AI on security posture tasks. OSB addresses the environment data gap by surfacing curated, synthetic-organization environments as a shared snapshot of the security state, and evaluating posture investigation across two modalities: text-to-SQL over a relational snapshot and each vendor's native API over a served instance of the same environment. The design freezes the target once and grades answers against a known-correct reference. Around this commitment it assembles the five components of Section~\ref{sec:components}, from the data layer through the bring-your-own path, with a scoring layer that reports each criterion on its own axis. We instantiated the framework with two identity-security packs and a family of environment datasets spanning multiple scales, and charted its extension along the surface, subdomain, and automation axes. Our aim is not a single leaderboard but a shared substrate: one on which the community can author packs, compare agents on the work they are deployed to do, and turn every evaluation into data that advances the next agent. On such a substrate, trusting a posture agent becomes a matter of measurement rather than judgment, a foundation for extending trustworthy agentic AI to the situational-awareness core of autonomous enterprise cyber defense.

\bibliographystyle{unsrt}
\bibliography{references}

\end{document}